\documentclass[twocolumn,letter]{jpsj3} 
\title{Contrasting Pressure Effects in Sr$_2$VFeAsO$_3$ and Sr$_2$ScFePO$_3$}

\author{Hisashi \textsc{Kotegawa}$^{1,4}$\thanks{E-mail address: kotegawa@crystal.kobe-u.ac.jp}, Takayuki \textsc{Kawazoe}$^{1}$, Hideki \textsc{Tou}$^{1,4}$, Keizo \textsc{Murata}$^{2}$, Hiraku \textsc{Ogino}$^{3,4}$, Kohji \textsc{Kishio}$^{3,4}$, and Jun-ichi \textsc{Shimoyama}$^{3,4}$}

\inst{$^{1}$Department of Physics, Kobe University, Kobe 657-8501\\
$^{2}$Division of Molecular Materials Science, Graduate School of Science, Osaka City University, Osaka 558-8585\\
$^{3}$Department of Applied Chemistry, The University of Tokyo, Tokyo 113-8656\\
$^{4}$JST, Transformative Research - Project on Iron Pnictides (TRIP), Chiyoda, Tokyo 102-0075\\
}

\abst{
We report the resistivity measurements under pressure of two Fe-based superconductors with a thick perovskite oxide layer, Sr$_2$VFeAsO$_3$ and Sr$_2$ScFePO$_3$. The superconducting transition temperature $T_c$ of Sr$_2$VFeAsO$_3$ markedly increases with increasing pressure. Its onset value, which was $T_c^{onset}=36.4$ K at ambient pressure, increases to $T_c^{onset}=46.0$ K at $\sim4$ GPa, ensuring the potential of the "21113" system as a high-$T_c$ material. However, the superconductivity of Sr$_2$ScFePO$_3$ is strongly suppressed under pressure. The $T_c^{onset}$ of $\sim16$ K decreases to $\sim5$ K at $\sim4$ GPa, and the zero-resistance state is almost lost.  We discuss the factor that induces this contrasting pressure effect.

}

\kword{Sr$_2$VFeAsO$_3$, Sr$_2$ScFePO$_3$, superconductivity, pressure}

\begin{document}
\maketitle

Intensive researches of the Fe-based superconductors and related materials have continued to date. 
In these researches, one of interests is in the relationship between superconducting transition temperature ($T_c$) and crystal structure.
Lee {\it et al.} have found the relationship between $T_c$ and the bond angle of As-Fe-As, and suggested that the formation of a regular FeAs$_4$ tetrahedron is important for higher $T_c$.\cite{Lee}
Meanwhile, Kuroki {\it et al.} have theoretically suggested that the pnictogen height from the Fe plane ($h_{Pn}$) and lattice constant are good parameters to control $T_c$.\cite{Kuroki}
They pointed out that the nesting properties of Fermi surfaces are sensitive to $h_{Pn}$ and, that the reduction in lattice constant suppresses superconductivity through the enhancement of the hopping between orbitals, and explained the difference in $T_c$ among some "1111" systems such as LaFeAs(O$_{1-x}$,F$_x$), NdFeAsO$_{1-y}$, LaFePO, and NdFeAsO$_{1-y}$ under pressure.
FeSe ("11" system) was also found to show a close relationship between $T_c$ and crystal structure, especially $h_{Pn}$.
The $T_c$ of FeSe, which is 8 K at ambient pressure, increases markedly under pressure,\cite{Mizuguchi,Margadonna2,Medvedev,Imai,Masaki} accompanied by a decrease in $h_{Pn}$.\cite{Margadonna2}
The pressure dependence of $T_c$ exhibits a plateau at approximately 1 GPa.\cite{Masaki}
$h_{Pn}$ shows a similar pressure dependence to $T_c$, and reproduces the plateau, while the bond angle of Se-Fe-Se is almost unchanged under pressure.\cite{Margadonna2}
NMR studies of FeSe have revealed that antiferromagnetic spin fluctuations develop under pressure associated with the increase in $T_c$.\cite{Imai,Masaki}
Results of these studies support that $h_{Pn}$ is a key factor for adjusting the nesting properties and $T_c$.
Between LaFePO ($h_{Pn}=1.12$ \AA) and NdFeAsO$_{1-y}$ ($h_{Pn}=1.38$ \AA),\cite{Kamihara2,Lee} $T_c$ increases with increasing $h_{Pn}$, while the $T_c$ of FeSe increases with decreasing $h_{Pn}$ ($h_{Pn}=1.45 \to 1.42$ \AA).\cite{Margadonna2}
This suggests that an optimum $h_{Pn}$ exists at approximately $h_{Pn}$ of NdFeAsO$_{1-y}$.

Recently, a new series of Fe-based superconductors with a thick perovskite oxide layer were discovered.\cite{Ogino,Zhu}
Ogino {\it et al.} have reported a superconductivity of 17 K in Sr$_2$ScFePO$_3$, which is the highest $T_c$ among those of FeP systems.\cite{Ogino}
On the other hand, Zhu {\it et al.} have reported a superconductivity of 37 K  in Sr$_2$VFeAsO$_3$.\cite{Zhu}
These compounds have a highly two-dimensional crystal structure from which good nesting properties are expected.
Another feature related to crystal structure is that the distance between Fe and pnictogen, which is closely connected to $h_{Pn}$, tends to be longer than that of "1111" systems.\cite{Ogino2}
Actually, the $h_{Pn}$ of Sr$_2$ScFePO$_3$ is estimated to be 1.20 \AA\, which is higher than that (1.12 \AA) of LaFePO.\cite{Ogino,Kamihara2}
In the case of Sr$_2$VFeAsO$_3$, $h_{Pn}$ is estimated to be 1.42 \AA\, which is higher than that (1.38 \AA) of NdFeAsO and close to that of FeSe under pressure.\cite{Zhu,Lee,Margadonna2}
In "11" systems like FeSe, there are no materials with a low $h_{Pn}$.
However, "21113" systems include materials with both high $h_{Pn}$ and low $h_{Pn}$.
It is expected that the pressure effect for these "21113" systems will give a clue to understanding the crucial factor for higher $T_c$.

Polycrystalline samples were synthesized by solid-state reaction.\cite{Ogino}
Sr$_2$ScFePO$_3$ has a low bulk density of less than 40\%.
Electrical resistivity ($\rho$) measurement at high pressures was carried out using an indenter cell.\cite{indenter}
Electrical resistivity was measured by a four-probe method using silver paste for contact.
The typical sample dimensions were $0.8\times0.3\times0.2$ mm$^3$.
We used Daphne 7474 for the resistivity measurement as a pressure-transmitting medium.\cite{Murata}
Applied pressure was estimated from the $T_{c}$ of the lead manometer.

\begin{figure}[htb]
\centering
\includegraphics[width=0.75\linewidth]{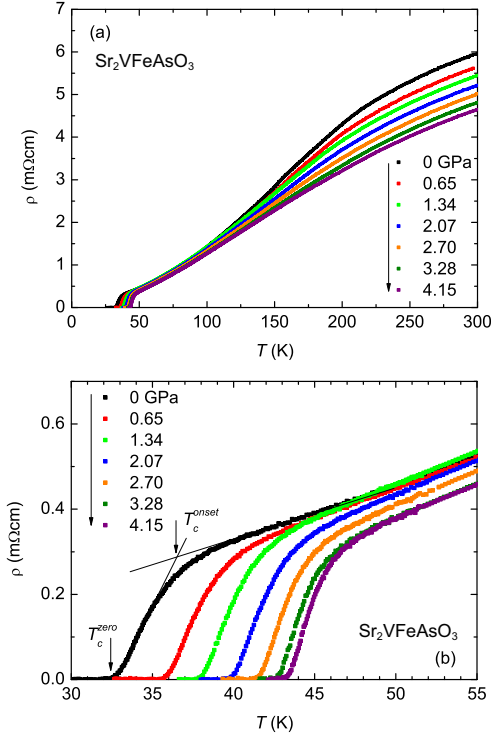}
\caption[]{(color online) Temperature dependences of $\rho$ in the $T$ ranges of (a) $0-300$ K and (b) $30-55$ K for Sr$_2$VFeAsO$_3$. $T_c$ increases with increasing pressure from $T_c^{onset}=36.4$ K and $T_c^{zero}=32.5$ K at ambient pressure to $T_c^{onset}=46.0$ K and $T_c^{zero}=43.2$ K at 4.15 GPa, respectively.
}
\end{figure}

Figures 1(a) and 1(b) show the temperature dependences of $\rho$ for Sr$_2$VFeAsO$_3$ at high pressures up to 4.15 GPa.
The onset temperature of the superconducting transition ($T_c^{onset}$) and the temperature of the zero resistance state ($T_c^{zero}$) are defined in the figure.
$T_c^{onset}=36.4$ K and $T_c^{zero}=32.5$ K at ambient pressure are comparable to those in the first report.\cite{Zhu}
These $T_c$'s increase with increasing pressure and reach $T_c^{onset}=46.0$ K and $T_c^{zero}=43.2$ K at 4.15 GPa.
The temperature dependence of $\rho$ above $T_c$ approaches a $T$ linear-like behavior from a $T^2$-like behavior.
The power $n$ in $\rho(T)=\rho_0+AT^n$ between 50 and 100 K is tentatively estimated to be $\sim1.8$ at ambient pressure and $\sim1.4$ at 4.15 GPa while treating $\rho_0$, $A$, and $n$ as fitting parameters, although $\rho_0$ becomes negative above $\sim2$ GPa in this fitting.

\begin{figure}[b]
\centering
\includegraphics[width=0.8\linewidth]{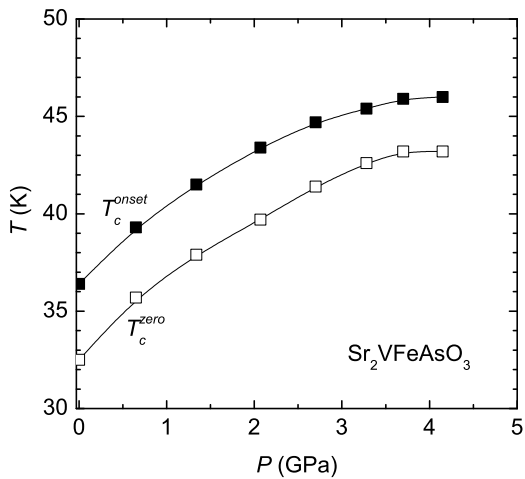}
\caption[]{Pressure dependences of $T_c^{onset}$ and $T_c^{zero}$ for Sr$_2$VFeAsO$_3$. $T_c$ first increases with a large slope of $dT_c/dP = 4.6$ K/GPa, which seems to saturate at approximately 4 GPa.
}
\end{figure}

Figure 2 shows the pressure dependences of $T_c^{onset}$ and $T_c^{zero}$ for Sr$_2$VFeAsO$_3$.
$T_c$ first increases with a large slope of $dT_c/dP = 4.6$ K/GPa, which seems to saturate at approximately 4 GPa.
$T_c$ increases monotonically up to 4.15 GPa without the plateau observed in FeSe.\cite{Masaki}
The $T_c$ of Sr$_2$VFeAsO$_3$ under pressure exceeds those of "122" systems such as (Ba,K)Fe$_2$As$_2$ and SrFe$_2$As$_2$ under pressure,\cite{Rotter,Kotegawa} but is lower than the highest $T_c$ of more than 50 K in "1111" systems.\cite{Ren,Kito}
Carrier density is also important for the mechanism of superconductivity, but it is unknown whether the carrier density of Sr$_2$VFeAsO$_3$ is optimum.
If we can adjust carrier density, a higher $T_c$ might be realized in "21113" systems.

\begin{figure}[htb]
\centering
\includegraphics[width=0.75\linewidth]{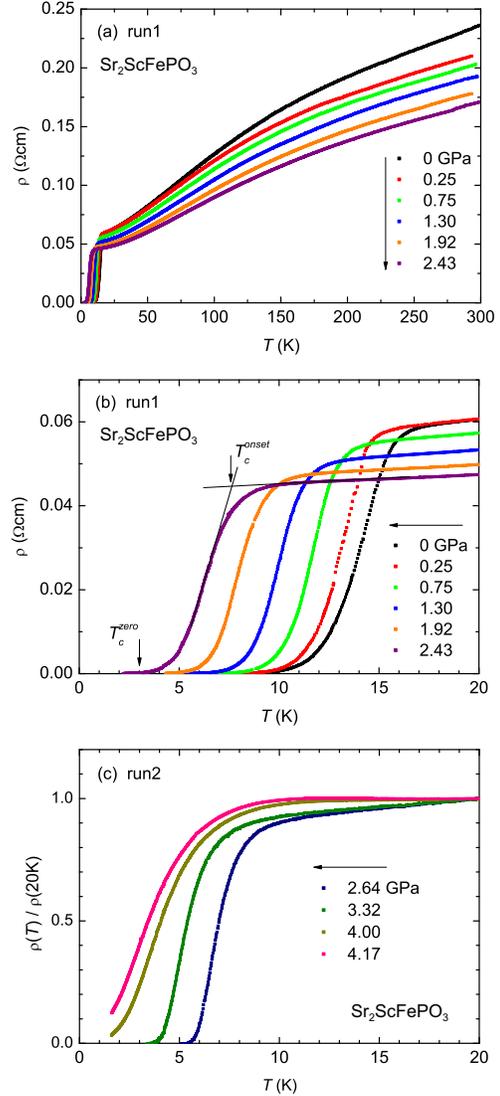}
\caption[]{(color online) Temperature dependences of $\rho$ below (a) $300$ and (b) $20$ K for Sr$_2$VFeAsO$_3$. Figure (c) shows the $\rho$ is measured for the different parts of the same sample. $T_c$ decreases with increasing pressure, and the zero-resistance state disappears at high pressures.
}
\end{figure}

Figures 3(a)-3(c) show the temperature dependences of $\rho$ for Sr$_2$ScFePO$_3$ at high pressures up to 4.17 GPa.
At ambient pressure, we estimated $T_c^{onset}=15.6$ K and $T_c^{zero}=8.4$ K.
This system exhibits a large superconducting transition width probably owing to the low bulk density of the sample.
We should note that a clear diamagnetic signal is observed below $15-17$ K.\cite{Ogino}
The transition width was large in run 1; it was sharp in run 2 using a different part of the same sample.
Since the absolute value of $\rho$ changed unnaturally in run 2, $\rho$ is normalized at 20 K.
As seen in the figure, the $T_c$ of this compound decreases significantly under pressure, and the zero-resistance state goes beyond the observed temperature range above 4 GPa.

\begin{figure}[htb]
\centering
\includegraphics[width=0.8\linewidth]{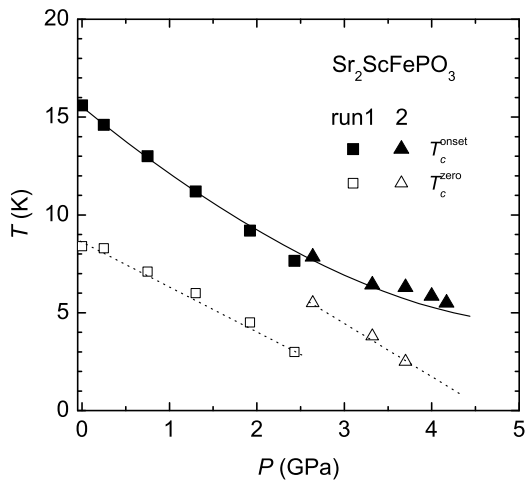}
\caption[]{Pressure dependences of $T_c^{onset}$ and $T_c^{zero}$ for Sr$_2$ScFePO$_3$. $T_c$ decreases with increasing pressure.
}
\end{figure}

Figure 4 shows the pressure dependences of $T_c^{onset}$ and $T_c^{zero}$ for Sr$_2$ScFePO$_3$.
$T_c^{onset}$ shows a large negative slope of $-3.6$ K/GPa at lower pressures and seems to saturate at high pressures.
$T_c^{zero}$ exhibits a strong sample-dependence, but approaches zero at approximately 4 GPa.

The $T_c$'s of Fe-based superconductors have been pointed out to be associated with their crystal structures.\cite{Lee,Kuroki}
Lee {\it et al.} showed a clear relationship between $T_c$ and the bond angle of As-Fe-As in various Fe-based superconductors, and suggested that a regular FeAs$_4$ tetrahedron is important for a high $T_c$. \cite{Lee}
However, although the $T_c$ of NdFeAsO$_{1-y}$ decreases monotonically under pressure,\cite{Takeshita} the pressure dependence of the bond angle of As-Fe-As seems to be not associated with that of $T_c$.\cite{Kumai}
Instead, Kuroki {\it et al.} have theoretically explained the reduction in $T_c$ under pressure in NdFeAsO$_{1-y}$ using the changes in $h_{Pn}$ and lattice constant.\cite{Kuroki}
They have pointed out that the nesting property is sensitive to $h_{Pn}$.
Also in FeSe, the pressure dependence of $T_c$ with a plateau at approximately 1 GPa is not associated with that of the bond angle of Se-Fe-Se but rather with that of $h_{Pn}$.\cite{Masaki,Margadonna2}
The FeSe$_4$ tetrahedron slightly deviates from the regular one, but $T_c$ rapidly increases.\cite{Margadonna2}
Thus, we consider that $h_{Pn}$ is a good parameter for explaining the pressure dependence of $T_c$.

The $T_c$, $h_{Pn}$, $h_{Pn}$ under pressure, and the pressure effects on $T_c$ for some Fe-based superconductors are summarized in Table 1.
The relationship between $T_c$ and $h_{Pn}$ is seen at ambient pressure.
Table~1 implies that the $h_{Pn}$ of NdFeAsO$_{1-y}$ is optimum for a high $T_c$.
Kuroki {\it et al.}, have suggested that lattice constant also affects superconductivity, and its reduction lowers $T_c$.
Lattice constant, which always decreases under pressure, cannot be a factor for the increase in $T_c$ under pressure.
Here, we take notice of $h_{Pn}$ and its change under pressure while considering that an optimum $h_{Pn}$ exists.
Note that Kuroki {\it et al.} did not mention that an optimum $h_{Pn}$ exists.\cite{Kuroki}
The structural parameters under pressure for FeSe, NdFeAsO$_{1-y}$, and LiFeAs have already been investigated.\cite{Margadonna2,Kumai}
In NdFeAsO$_{1-y}$, $h_{Pn}$ slightly decreases from $1.34$ to $1.33$ \AA\ under pressure.\cite{Kumai}
Note that this $h_{Pn}$ at ambient pressure differs from that (1.38 \AA) reported by Lee {\it et al.}\cite{Lee}
The decrease in $h_{Pn}$ and the reduction in lattice constant are considered to suppress superconductivity under pressure in NdFeAsO$_{1-y}$.\cite{Kuroki}
If we consider that the $h_{Pn} =1.34-1.38$ \AA\ of NdFeAsO$_{1-y}$ is the optimum range for a high $T_c$, the $h_{Pn}$ of FeSe approaches it under pressure, and $T_c$ increases.
In contrast, the $h_{Pn}$ of LiFeAs deviates from the optimum range, and $T_c$ decreases.\cite{Mito,Gooch}
In the case of LiFeAs, a FeAs$_4$ tetrahedron also deviates from the regular one under pressure.\cite{Mito}
The change in $h_{Pn}$ is likely the main factor in deciding the change in $T_c$ under pressure in NdFeAsO$_{1-y}$, FeSe, and LiFeAs.
In this context, if the $h_{Pn}$ in "21113" systems decreases under high pressures, the increase in $T_c$ under pressure for Sr$_2$VFeAsO$_3$ is reasonable because the $h_{Pn}$ is high at ambient pressure.
Interestingly, the $T_c$ and $h_{Pn}$ of Sr$_2$VFeAsO$_3$ are comparable to those of FeSe under pressure.
On the other hand, the decrease in $T_c$ under pressure for Sr$_2$ScFePO$_3$ is also reasonable because the $h_{Pn}$ at ambient pressure is already low.

However, we should carefully study other factors that change under pressure, such as the carrier density and the magnetism of Fe or other magnetic ions, in discussing $T_c$ under pressure.
The pressure dependence of $h_{Pn}$ also has to be determined experimentally.
The $T_c$'s of LaFeAs(O$_{1-x}$F$_x$) and LaFePO increase under pressure.\cite{Takahashi,Hamlin,Igawa}
If the $h_{Pn}$ of these compounds decreases under pressure as well as that of NdFeAsO$_{1-y}$, which is also a "1111" system, the increase in $T_c$ cannot be explained.
To our knowledge, there is as yet no report concerning the structural parameters of these compounds under pressure.
It is interesting how the $h_{Pn}$'s of these compounds change under pressure.
The pressure dependences of $T_c$ for hole-doped K$_x$Sr$_{1-x}$Fe$_2$As$_2$ systems have been investigated.\cite{Gooch2}
The $T_c$ of a lightly-doped system increases under pressure, while the $T_c$ of a heavily-doped system decreases.
It is doubtful whether these pressure effects can be explained by only the change in $h_{Pn}$.

Another aspect that should be mentioned is the difference in the band calculation between Sr$_2$VFeAsO$_3$ and Sr$_2$ScFePO$_3$.\cite{Shein1,Shein2}
Although the Sr$_2$VO$_3$ layer in Sr$_2$VFeAsO$_3$ is metallic, the Sr$_2$ScO$_3$ layer is insulating.\cite{Shein1,Shein2}
The replacement of Sc with V is reported to lead to a drastic change in the Fermi surface.
The existence of the same nesting properties as those of other Fe-based superconductors seems to be controversial in Sr$_2$VFeAsO$_3$.\cite{KWLee,Mazin}
We may have to carefully determine whether the superconductivity in Sr$_2$VFeAsO$_3$ is understood in the same framework as those in other Fe-based superconductors.

In summary, we have investigated the pressure effects of Sr$_2$VFeAsO$_3$ and Sr$_2$ScFePO$_3$ through resistivity measurement.
The $T_c$ of Sr$_2$VFeAsO$_3$ significantly increases, while the $T_c$ of Sr$_2$ScFePO$_3$ decreases under pressure.
We conjecture that these opposite pressure effects mainly originate from the difference in $h_{Pn}$ at ambient pressure.
The $T_c^{onset}$ of Sr$_2$VFeAsO$_3$ reaches 46 K at $\sim4$ GPa, and exceeds the maximum $T_c$ of "122" and "11" systems.
The present results are expected to give a clue to obtaining a higher $T_c$ in "21113" systems.

We thank S. Sato and Y. Shimizu for cooperation in the sample preparation, K. Ishida for helpful discussions, and S. Clarke for useful information.
This work has been partly supported by Grants-in-Aid for Scientific Research (Nos. 19105006, 20740197, and 20102005) from the Ministry of Education, Culture, Sports, Science, and Technology (MEXT) of Japan.

\clearpage

\begin{table}[ht]

 \begin{center}
     \caption{$T_c$, $h_{Pn}$, $h_{Pn}$ under pressure, and pressure effects on $T_c$ (maximum $T_c^{onset}$ under pressure) for some Fe-based superconductors. For LiFeAs, FeSe, and NdFeAsO$_{1-y}$, the $h_{Pn}$'s under pressure have been determined. The relationship between $T_c$ and $h_{Pn}$ is observed at ambient pressure and under pressure, suggesting that the $h_{Pn}$ range of $1.34-1.38$ \AA\ of NdFeAsO$_{1-y}$ is optimum for a high $T_c$. If $h_{Pn}$ decreases under pressure in "21113" systems, the pressure effects in Sr$_2$VFeAsO$_3$ and Sr$_2$ScFePO$_3$ are reasonable. LaFeAs(O$_{1-x}$F$_x$) and LaFePO may be exceptions, if $h_{Pn}$ decreases under pressure as well as NdFeAsO$_{1-y}$. The pressure effect of K$_{x}$Sr$_{1-x}$Fe$_2$As$_2$ may also be difficult to be explained by only the change in $h_{Pn}$. $h_{Pn}$ is estimated to be 1.36 for $x=0$ and 1.43 for $x=1$ at ambient pressure.\cite{Tegel,Rotter2}
    }
  \begin{tabular}{lclll}

        & $T_c$(K) & $h_{Pn}$(\AA) &  $h_{Pn}$ under pressure (\AA) & $T_c$ under $P$     \\
   \hline
                               &    &  &                  &         \\
LiFeAs                         & 18 & 1.50 \cite{Pitcher} & $1.55 \to 1.56$ ($\sim4$ GPa) \cite{Mito}           & down \cite{Mito,Gooch}   \\
FeSe                           & 8  & 1.46 \cite{Margadonna} & $1.45 \to 1.42$ ($\sim5$ GPa) \cite{Margadonna2} & up, 37 K \cite{Mizuguchi,Margadonna2,Medvedev,Masaki}   \\
Sr$_2$VFeAsO$_3$ & 36 & 1.42 \cite{Zhu}  &          & up, 46 K   \\
NdFeAsO$_{1-y}$                & 54 & 1.38 \cite{Lee} & $1.34 \to 1.33 $ (7.5 GPa)\cite{Kumai} & down \cite{Takeshita}   \\
LaFeAs(O$_{1-x}$F$_x$)         & 26 & 1.32 \cite{Kamihara}  &          & up, 43 K \cite{Takahashi}   \\
Sr$_2$ScFePO$_3$ & 16 & 1.20 \cite{Ogino}      &      & down   \\
LaFePO                         & 7  & 1.12 \cite{Kamihara2}  &          & up, 10 K \cite{Hamlin,Igawa}   \\
                               &    &                    &         \\ 
K$_{0.2}$Sr$_{0.8}$Fe$_2$As$_2$   & 13  &      &      & up, 16 K \cite{Gooch2}       \\
K$_{0.4}$Sr$_{0.6}$Fe$_2$As$_2$   & 37  &  &          & constant \cite{Gooch2}      \\
K$_{0.7}$Sr$_{0.3}$Fe$_2$As$_2$   & 8  &     &        & down \cite{Gooch2}      \\
                               &    &         &           &         \\ 
   \hline
  \end{tabular}
     \end{center}
\end{table}

\end{document}